\documentstyle[epsfig,12pt]{article}
\newcommand{\tch}{{t\rightarrow c\,h}}
\newcommand{\sbt}{{\tilde b}} 
\newcommand{\st}{{\tilde t}}
\newcommand{\charg}{{\tilde \chi^+}}

\newcommand{\slas}[1]{\rlap/ #1}
\newcommand{\pl}{P_L} 
\newcommand{\pr}{P_R}
\newcommand{\tb}{{\tan\beta}} 
\newcommand{\GeV}{{\rm\,GeV}}
\newcommand{\stackm}{\stackrel{\scriptstyle <}{{ }_{\sim}}}

\begin{document}

\def\pubnum{427}
\def\data{October, 1997}

\def\UUAABB{\hbox{
    \vrule height0pt width5in
    \vbox{\hbox{\rm 
     UAB-FT-\pubnum
    }\break\hbox{\data\hfill}
    \break\hbox{hep-ph/9710267\hfill} 
    \hrule height2.7cm width0pt}
   }}   
\UUAABB
\vspace{3cm}
\begin{center}
  \begin{large}
    \begin{bf}
      FCNC top decays into Higgs bosons in the MSSM\footnote{Talk presented at
        the {\em International Workshop on quantum effects in the MSSM},
        Barcelona, 9-13 September, 1997. To appear in the proceedings, World
        Scientific, ed. J. Sol{\`a}.}
      \\
    \end{bf}
  \end{large}
  \vspace{1cm}
  Jaume GUASCH\\

  \vspace{0.25cm} 
  Grup de F{\'\i}sica Te{\`o}rica\\ 
  and\\ 
  Institut de F{\'\i}sica d'Altes Energies\\ 
  \vspace{0.25cm} 
  Universitat Aut{\`o}noma de Barcelona\\
  08193 Bellaterra (Barcelona), Catalonia, Spain\\
\end{center}
\vspace{0.3cm}

\begin{abstract}
  We compute the partial width of the FCNC top quark decay $t\rightarrow c\,h$
  in the framework of the 
  Minimal Supersymmetric Standard Model, where $h\equiv h^0,H^0,A^0$ is any of
  the neutral Higgs of the MSSM. We include the SUSY electroweak, Higgs, and
  SUSY-QCD contributions. Our results substantially improve previous estimations
  on the subject, and we find that there is a possibility that they can be
  measured at {\bf LHC}.
\end{abstract}

\newpage
\section{Introduction}

We perform the computation of the Flavour Changing Neutral Current
(FCNC) decay of the top quark into a charm quark and a neutral Higgs particle in
the framework of the 
Minimal Supersymmetric Standard Model (MSSM), $\tch$ where $h$ is any
of the neutral Higgs particles of the MSSM, namely the light or heavy CP-even
(``scalar'') Higgs 
bosons ($h^0$, $H^0$), or the CP-odd (``pseudoscalar'') Higgs ($A^0$)\footnote{We use a
  widely known notation\cite{Hunter}; see Refs.\cite{CGGJS}  for further
  details and conventions.}. We compute the contributions from the SUSY 
electroweak, Higgs, and SUSY-QCD sectors, in a sparticle mass model motivated by model
building and Renormalization Group Equations (RGE). However, we neither restrict
ourselves
to a spectrum of any SUSY-GUT model (such as SUGRA) --which would constrain
the masses in a narrow range--, nor to a generic, phenomenological motivated,
spectrum --which would have too many parameters to play with.

There exist some computations of FCNC top quark decays, both in the SM and in
the MSSM\cite{FCNCSM,tcv,tch}. There has
been some work concerning the decay channel into gauge bosons ($t\rightarrow
c\,V$, $V\equiv \gamma,Z,g$), see for example Ref.\cite{tcv} for some works on
the subject. The conclusion of these works is that the branching ratio of this
decay is at most $10^{-5}$, maybe a bit larger in the gluon channel. However,
to our knowledge, there are not so many works on the FCNC decay of the top quark
into Higgs in the MSSM\cite{tch}, and they are not so complete as in the case
of the gauge bosons. For example in~\cite{tch} it is concluded that the
branching ratio for 
the decay channel
$\tch$ in the MSSM is at most of $10^{-9}$, for the SUSY electroweak
contributions, and 
$10^{-5}$, for the SUSY-QCD contributions. However we think that the work
of~\cite{tch} is not complete. They do not include effects of the Higgs particles
in the loops, and they do not take into account the $\tilde q_L\,\tilde q_R\,h$
vertices, so they miss the potentially large contributions coming from the
trilinear soft-SUSY-breaking terms $A_{t,b}$, and from the higgsino mass
parameter $\mu$. We find that a full treatment
of the SUSY-QCD contributions may greatly enhance the FCNC width by some orders
of magnitude. Therefore, a more general, and more rigorous, computation of the
decay $\tch$ is mandatory\cite{GStch}. 

In section~\ref{sec:gen} we make a summary of the technics of the
computation, in sections~\ref{sec:ew} and~\ref{sec:qcd} we present our results
for the SUSY electroweak and the SUSY-QCD contributions to the decay width $\tch$
respectively, and finally we present the conclusions.

\section{Relevant lagrangians and form factors}
\label{sec:gen}

The computation of FCNC processes at one loop, unlike other calculations,
does not involve renormalization of parameters, or wave functions, so one is
left only with the computation of the different diagrams that contribute to the
process. The generic one loop diagrams contributing to the decay under study are
in fig.~\ref{fig:diagrams}. The vertex diagram $V$ follows after a straightforward
calculation. As for the diagrams $S_t$ and $S_c$ we define a mixed self-energy,
\begin{equation}
  \label{eq:automixta}
   \Sigma_{tc}(k) \equiv
  \slas{k}\,\Sigma_L(k^2)\,\pl+\slas{k}\,\Sigma_R(k^2)\,\pr+ 
  m_t\,\left(\Sigma_{Ls}(k^2)\,\pl+\Sigma_{Rs}(k^2)\,\pr\right)
\end{equation}
--where the $m_t$ factor multiplying the scalar part is arbitrary, 
put there only to maintain the same units between the different $\Sigma_i$--, and compute
the effects of $\Sigma_i$ to the amplitude. We compute the different contributions
to $V$, $S_t$ and $S_c$ and define an 
``effective'' vertex
\begin{equation}
\label{eq:effvertex}
 -i\,T\equiv -i\,g\, \bar{u}_c(p)\, \left( F_L\,P_L+ F_R\,P_R\right)\,
  u_t(k)\,\,\,.
\end{equation}

We have taken into account all three generations of quarks and squarks, and have
performed the usual checks of
the computation, in particular that the form factors $F_L$ and $F_R$  are free
of divergences before adding up 
the three quark generations, both analytically and numerically in the
implementation of the code.

After squaring the matrix element~(\ref{eq:effvertex}), and multiplying by the
phase space 
factor, we can compute the decay width, and define the ratio
\begin{equation}
\label{eq:defbr}
 B(\tch)\equiv\frac{\Gamma
          (t\rightarrow c\,h)}{\Gamma(t \rightarrow b\,W^+)}
\end{equation}
which will be the main object under study. This ratio is not the total branching
fraction of this decay mode, as there are many other channels that should be
added up to the denominator of (\ref{eq:defbr}) in the MSSM, such as the two and
three body 
decays of the top quark into SUSY particles, and also the decay channel $t\rightarrow
H^+\,b$\cite{CGGJS,talksola}. For the mass spectrum used in the numerical
analysis (see sections~\ref{sec:ew} and~\ref{sec:qcd}) the former are phase
space closed, whereas the latter could have a large branching ratio.

We would like to single out two pieces of the interaction lagrangian, namely the
ones involving higgsino-sbottom-charm and Higgs-bottom-charm
\begin{eqnarray}
  \label{eq:mainew}
  {\cal L}_{\tilde h\,\tilde b\,c}&=&- g\,V_{cb}\,\bar{c} \left({R_{1a}\,\lambda_c}\, P_L+
    {R_{2a}\,\lambda_b}\, P_R \right) \chi^+ \tilde{b}_a \nonumber\\
  {\cal L}_{H\,b\,c}&=&\frac{g}{\sqrt{2}M_W}
  V_{cb}\,\bar{c}\left({m_c\,\cot\beta}\,P_L+{m_b\,\tan\beta}\, P_R\right)\,b\,H^+\,\,\,,
\end{eqnarray}
where $V_{cb}$ is the CKM matrix element, and $\lambda_{c,b}$ are the charm and
bottom Yukawa couplings\footnote{We refer again to~\cite{CGGJS} for conventions
  and notation.}. Looking at these two pieces of the lagrangian one can get an
estimation of the relative importance of the different form
factors~(\ref{eq:effvertex}) (see section~\ref{sec:ew}).

\section{SUSY-EW contributions}
\label{sec:ew}

For the electroweak contributions to the decay channel $\tch$ we work in the so
called Super-CKM basis, that is, we take the
simplification that the squark mass matrix diagonalizes as the quark mass
matrix, so that FCNC processes appear at one loop through the charged sector (charged
Higgs and charginos) with the same mixing matrix elements as in the Standard
Model (the CKM matrix). We introduce, as usual, the left-right mass matrix mixing
elements between
squarks\cite{CGGJS}: $m_q\,(A_q-\mu\{\tan\beta,\cot\beta\})$.

We have taken into account the contributions from charginos ($\charg_i$) and
down type squarks ($\tilde d_\alpha$, $\alpha=1,2,\ldots,6\equiv
d_1,d_2,\ldots,b_2$, the mass eigenstates down squarks), and from charged Higgs and
Goldstone bosons ($H^+,G^+$) and down type quarks ($d,s,b$). We have not
included the diagrams with gauge bosons ($W^+$) as the largest contributions
will come from the Yukawa couplings of the top and (at large $\tb$) bottom
quarks. However, the leading terms from longitudinal $W^+$ are included through
the inclusion of Goldstone bosons.

The input parameters chosen to illustrate the results in 
figs.~\ref{fig:formew}-\ref{fig:resultsew} are:
\begin{equation}
  \label{eq:inputew}
  \begin{array}{rcl}
    \tb&=&35 \\
    \mu&=&-100 \GeV \\
    M&=&150 \GeV\\ 
    M_{A^0}&=&60 \GeV\\ 
    m_{\tilde t_1}&=&150 \GeV\\ 
    m_{\tilde b_1}=m_{\tilde q}&=&200 \GeV\\
    A_t=A_q&=&300 \GeV\\
    A_b&=&-300 \GeV
  \end{array}
\end{equation}
where $m_{\st_1},m_{\sbt_1}$ are the lightest $\st$ and $\sbt$ mass, and all the
masses are above present experimental bounds\cite{limitssusy}-\cite{CDFlim}, 
though the mass of the
pseudoscalar Higgs ($M_{A^0}$) is in the edge of present {\bf LEP} bounds\cite{LEPma}, and
it is possibly excluded by the last analyses on the charged Higgs
mass\cite{talksola,CDFlim,GSlim,limitsmh}. However, this light Higgs mass is not essential in the
results, as can bee seen in fig.~\ref{fig:resultsew}~(d). We have chosen a SUSY mass spectrum around $200\GeV$, which
is not too light, so the results will not be artificially optimized. We have also checked all
through the numerical analysis that other bounds on experimental parameters
(such as $\delta\rho$) are fulfilled.

In fig.~\ref{fig:formew} we have plotted the different form factors
of~(\ref{eq:effvertex}) as a function of $\tb$ for the channel with the lightest
scalar Higgs ($h^0$). We can see that the contributions from the Higgs sector
and the contributions from the chargino sector are of the same order. It
turns out that they can be either of the same sign, or of opposite sign. The
chosen negative value for $A_b$ is to make the two contributions of the same
sign. It is also clear that in both cases $F_R \gg F_L$. This can be easily
understood by looking at the interaction lagrangians involving higgsino-sbottom-charm and
Higgs-bottom-charm~(\ref{eq:mainew}) 
where we can see that in both of them the contribution to the right-handed form factor
will be enhanced by the Yukawa coupling of the bottom quark, compared with the
charm Yukawa coupling that will contribute to the left-handed form factor. On the other
hand we have checked that the inclusion of the first two generations of quarks
and squarks only has an effect of a few percent on the total result.

In fig.~\ref{fig:resultsew} we can see the evolution of the
ratio~(\ref{eq:defbr})
with various parameters of the MSSM, by taking into account only the electroweak
contributions. The growing of the width with  $\tb$ 
(fig.~\ref{fig:resultsew}~(a)) shows that the bottom Yukawa coupling plays a
central role in these contributions. The evolution with the trilinear coupling
$A_b$ and the higgsino mass parameter $\mu$ --the two parameters that appear in
the trilinear coupling $\sbt_L\,\sbt_R\,h$-- displayed in figs.~\ref{fig:resultsew}~(b)
and~(c) shows that this parameters can 
enhance the width some orders of magnitude. We have artificially let $A_b$ grow up to
large scales (that are not allowed if one wants that squarks do not develop
vacuum expectation values) in order to emphasize the dependence on $A_b$. The various spikes
in these figures reflect the points where the form factors change sign, whereas
the shaded region in fig.~\ref{fig:resultsew}~(c) reflects the exclusion region of
$\mu$ by present {\bf LEP} bounds on the chargino mass. 

In all these figures the ratio~(\ref{eq:defbr}) is smaller for the heaviest
scalar Higgs ($H^0$) because with the parameters~(\ref{eq:inputew}) the CP-even
Higgs mixing angle $\alpha$ is near $-\pi/2$, so making the couplings of $H^0$ with down
quarks and squarks much weaker, but in fig.~\ref{fig:resultsew}~(d) it can be seen
that when the pseudoscalar Higgs mass grows (and this shifts $\alpha$ far away
from $-\pi/2$) the two scalar Higgs change roles.

We conclude that the typical value of the ratio~(\ref{eq:defbr}), at large
$\tan\beta\stackm 50$ and for a SUSY
spectrum around $200\GeV$, is
\begin{equation}
  \label{eq:conew}
  B^{\rm SUSY-EW}(\tch) \simeq 10^{-7}\,\,\,.
\end{equation}

This is an improvement of the previous results\cite{tch}, specially in the $A^0$
channel, by 2 orders of magnitude. 

\section{SUSY-QCD contributions}
\label{sec:qcd}

The gluino-mediated supersymmetric strong interactions in the MSSM can also produce FCNC
processes. This occurs when the squark 
mass matrix does not diagonalize with the same matrix as the one for the
quarks. We introduce then intergenerational mass terms for the squarks, but in
order to prevent the number of parameters from being too large, we have allowed (symmetric)
mixing mass terms only for the left-handed squarks. This simplification is often used in the
MSSM, and is justified by RGE analysis\cite{duncan}.

The mixing terms are introduced through the parameters $\delta_{ij}$ defined as
\begin{equation}
  \label{eq:defdelta}
(M^2_{LL})_{ij}=m_{ij}^2\equiv\delta_{ij}\,m_i\,m_j\,\,,
\end{equation}
where $m_i$ is the mass of the left-handed $i$ squark, and $m^2_{ij}$ is the mixing
mass matrix element between the generations $i$ and $j$. These parameters are
constrained by low energy data on FCNC\cite{gabbiani}. The bounds have been
computed using some approximations, so they must be taken as order of magnitude
limits. We use the following bounds on the $\delta$ parameters\cite{gabbiani}
\begin{eqnarray}
  \label{eq:limdelta}
  |\delta_{12}|&<&.1 \,\sqrt{m_{\tilde u}\,m_{\tilde c}}/{500 \GeV} \nonumber\\
  |\delta_{13}|&<&.098\,\sqrt{m_{\tilde u}\,m_{\tilde t}}/{500 \GeV}\nonumber\\
  |\delta_{23}|&<&8.2 \,m_{\tilde c}\,m_{\tilde t}/{(500 \GeV)^2}\,\,\,.
\end{eqnarray}

We compute the contributions to diagrams of fig.~\ref{fig:diagrams} with gluinos
($\tilde g$) and up squarks ($\tilde u_\alpha$, $\alpha=1,2,\ldots,6$). We use
the same input parameters as in the electroweak
contributions~(\ref{eq:inputew}), except for the $\mu$ parameter, namely
\begin{eqnarray}
  \label{eq:inputqcd}
  \mu&=&-200\GeV\nonumber\\
  m_{\tilde g}&=&150 \GeV\nonumber\\
  \delta&=&\left(\begin{array}{ccc}
      0 &0.03 &0.03\\
      &0&0.6\\
      &&0
    \end{array}\right)\,\,\,.
\end{eqnarray}

A comment is in order for the present set of inputs: we have introduced
in~(\ref{eq:inputew}) the lightest stop mass as an input, and this stop is mostly a
$\st_R$. However, in this new parametrization we introduce this mass as the lightest
$\tilde u_\alpha$ mass, which will be mostly a $\st_R$. On the other hand, we use
a larger absolute value for the $\mu$ parameter to enhance the ratio~(\ref{eq:defbr}).

Again the largest contribution comes from the right-handed form factor
of~(\ref{eq:effvertex}), but this is only because we have chosen not to
introduce mixing between right-handed squarks. 

We have plotted the evolution of the ratio~(\ref{eq:defbr}) with some parameters
of the MSSM in fig.~\ref{fig:resultsqcd}. As can be easily guessed, the most
important parameter for these contributions is the mixing mass parameter between
the 2nd and 3rd generation of left-handed squarks, the
less restricted one of the three (eq.~(\ref{eq:limdelta})). In
fig.~\ref{fig:resultsqcd}~(a) it is shown that changing $\delta_{23}$ by 3
orders of magnitude, the ratio~(\ref{eq:defbr}) can increase by 7 orders of
magnitude! We can see in fig.~\ref{fig:resultsqcd}~(b) that the $\mu$ parameter
also plays an important role, like in the electroweak contributions
(fig.~\ref{fig:resultsew}~(c)), and for the same reasons, bringing the
ratio~(\ref{eq:defbr}) up to values of $10^{-4}$. Notice that the central region
of $|\mu|\stackm 90\GeV$ is excluded by present {\bf LEP} bounds on the chargino
mass. 

The evolution with the
gluino mass (fig.~\ref{fig:resultsqcd}~(c)) is asymptotically quite stable, showing a
slow decoupling. Finally in
fig.~\ref{fig:resultsqcd}~(d) we have plotted the evolution with the pseudoscalar Higgs
mass, it is also quite stable, until near the kinematic limit for $A^0$ and
$H^0$.

We conclude that the typical value of the SUSY-QCD contributions
to~(\ref{eq:defbr}), with a SUSY spectrum around $200\GeV$, is
\begin{equation}
  \label{eq:conqcd}
    B^{\rm SUSY-QCD}(\tch) \simeq {\cal O}\,(10^{-5})\,\,\,,
\end{equation}
but in favourable regions of the parameter space (i.e. large $\mu$, or
relatively light gluino) it can easily reach values of $10^{-4}$. This is 1-2
orders of magnitude larger from the previous estimate\cite{tch}.

\section{Conclusions}
\label{sec:con}

We have computed the SUSY-electroweak, Higgs, and SUSY-QCD contributions to the
FCNC top quark decay $\tch$ ($h=h^0,H^0,A^0$) in the MSSM, using a mass spectrum
motivated, but not fully restricted, by model building and Renormalization Group
Equations.

We have found that with a SUSY mass spectrum around $200\GeV$, which is well above
present bounds\cite{limitssusy,LEPma,CDFlim}, the different contributions to this decay
are typically of the order 
\begin{eqnarray}
  \label{eq:confinal}
  B^{\rm SUSY-EW}(\tch) &\simeq& 10^{-7}\nonumber\\
  B^{\rm SUSY-QCD}(\tch)&\simeq& 10^{-5} - 10^{-4}\,\,\,.
\end{eqnarray}

The difference of at least two orders of magnitude between the two contributions
makes worthless to compute the 
interference between the two contributions, but if the limits on $\delta_{23}$
(eq.~(\ref{eq:limdelta})) improve, it should be necessary to make the full
computation.

The results~(\ref{eq:confinal}) are an improvement of the previous
results\cite{tch}, specially in the $A^0$ channel, thanks to the inclusion of
the $\tilde q_L\,\tilde q_R\,h$ vertex. 

It would probably be difficult that this decay can be measured either at the {\bf Tevatron},
or at the {\bf NLC}, but there exists a possibility for {\bf LHC}. As an example
to assess the
discovery reach of this 
accelerators the FCNC top quark decays into a vector boson are\cite{limits}
\begin{eqnarray}
  \label{eq:limtop}
  {\rm \bf LHC:}&B(t\rightarrow c\,V)& > 5\times 10^{-5} \nonumber\\
  {\rm \bf NLC:}&B(t\rightarrow c\,V)& > 10^{-3}-10^{-4} \,\,\,,
\end{eqnarray}
where the lack of sensitivity of {\bf NLC} is due to the lower luminosity. So,
if the discovery reach for FCNC Higgs processes are not very different from that
of the gauge bosons, there is a possibility to measure this decay channel at
{\bf LHC} even if SUSY particles are not seen at {\bf LEPII}.

{\bf Acknowledgements:} This talk is based in a work done in collaboration with
  Joan Sol{\`a}\cite{GStch}, whom I am very thankful for his interesting comments and
  discussions. This work has been supported by a grant 
  of the Comissionat per a Universitats i Recerca, Generalitat de Catalunya
  No. FI95-2125, and has also been partially financed by CICYT under project
  No. AEN95-0882.

\section*{Figure Captions}

\renewcommand{\labelenumi}{{\bf Fig.~\theenumi}}

\begin{enumerate}
\item Generic one loop Feynman Diagrams contributing to
  $\tch$\label{fig:diagrams}.
\item Different form factors (\ref{eq:effvertex}) for the
      channel $t\rightarrow c\,h^0$ as a 
      function of $\tb$, with the typical set of inputs of
      eq.(\ref{eq:inputew}).\label{fig:formew}
\item Evolution of the ratio (\ref{eq:defbr}) with (a) $\tb$, (b) the
    trilinear coupling $A_b$, (c) the higgsino mass parameter $\mu$, and (d) the
    pseudoscalar Higgs mass $M_{A^0}$, the rest of inputs are given in
    eq.(\ref{eq:inputew}).  \label{fig:resultsew}
\item  Evolution of the ratio (\ref{eq:defbr}) with (a) the mixing
    parameter between the 2nd and 3rd squark generations $\delta_{23}$, (b) the
    higgsino mass parameter $\mu$, (c) the gluino mass $m_{\tilde g}$, and (d)
    the 
    pseudoscalar Higgs mass $M_{A^0}$, the rest of inputs are given in
    eqs.(\protect{\ref{eq:inputew}}) and~(\protect{\ref{eq:inputqcd}}).\label{fig:resultsqcd}
\end{enumerate}

\renewcommand{\labelitemi}{\theenumi. }
\newpage
\pagestyle{empty}

\begin{center}
  \begin{tabular}{ccc}
    \epsfig{file=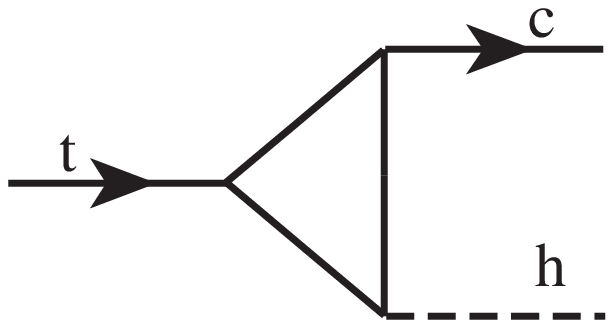,width=4.8cm} &
    \epsfig{file=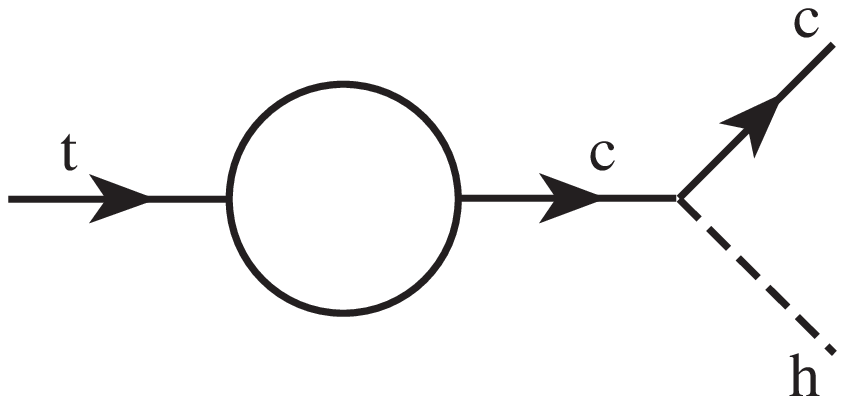,width=4.8cm} &
    \epsfig{file=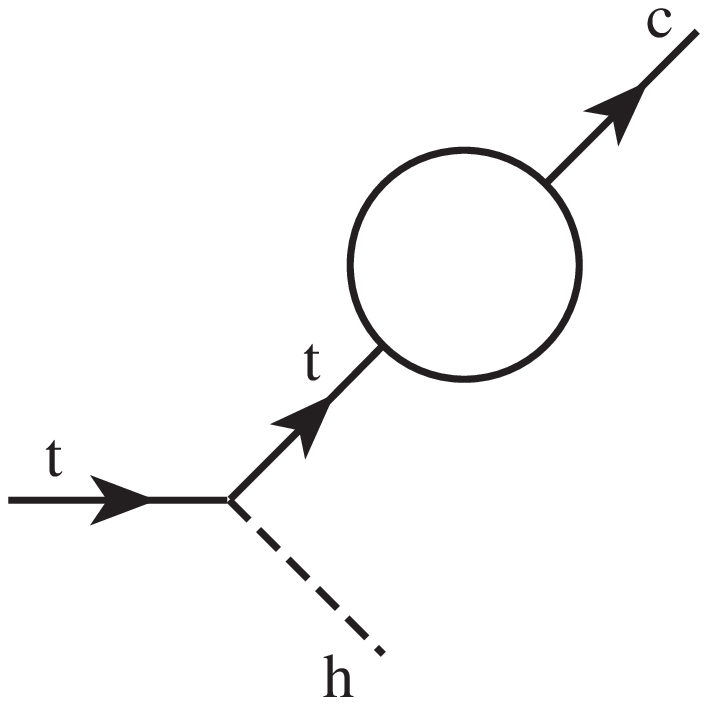,width=4.8cm} \\
    $V$ & $S_t$ & $S_c$
  \end{tabular}

\vspace{.3cm}

{\Large Fig. 1}
\end{center}

\vspace{2cm}

\begin{center}
  \epsfig{file=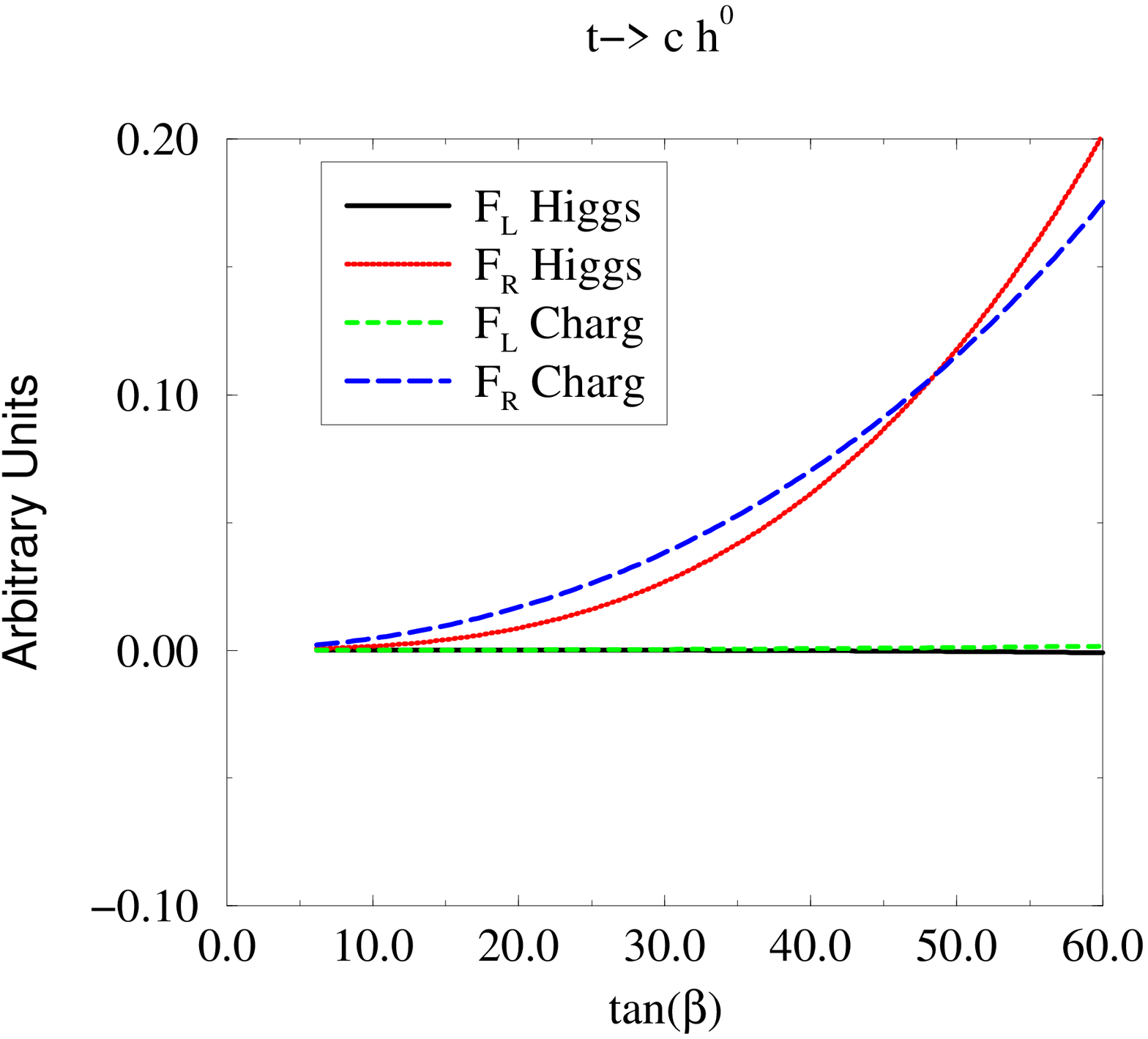,width=7cm}

\vspace{.3cm}

  {\Large Fig. 2}                      
\end{center}

\newpage

\begin{center}
  \begin{tabular}{cc}
    \epsfig{file=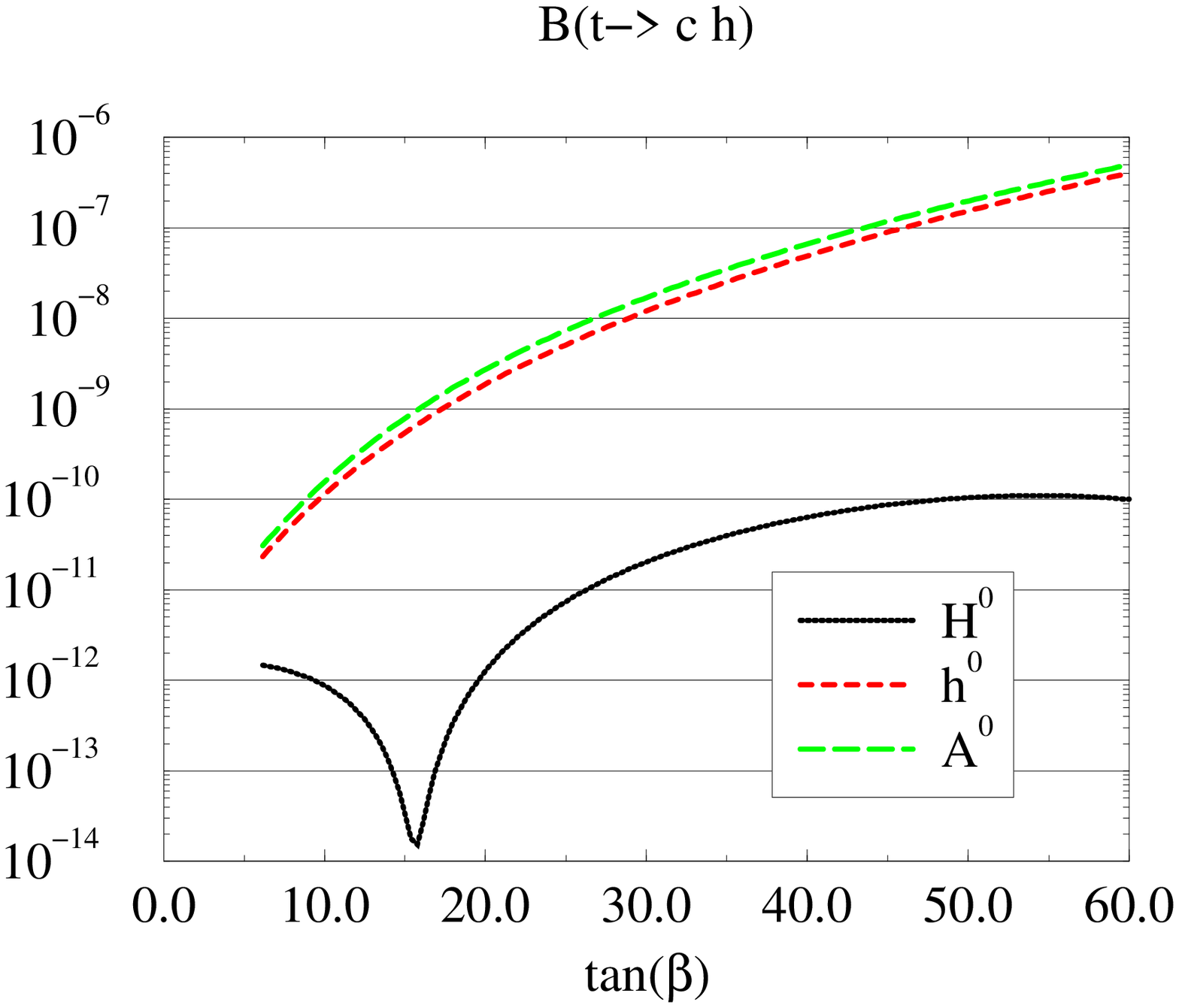,width=7cm} &
    \epsfig{file=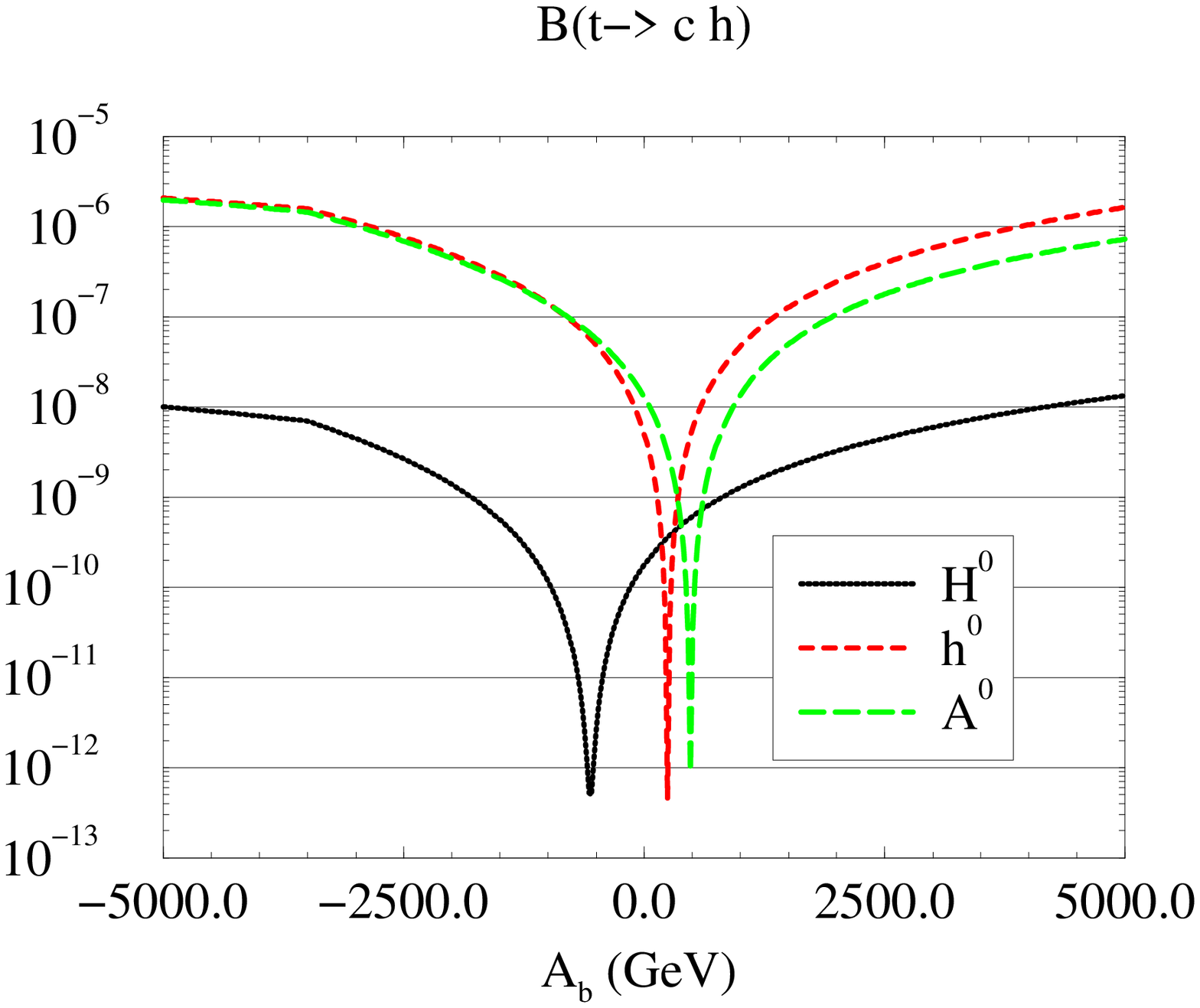,width=7cm}\\
    (a) & (b) \\
    \epsfig{file=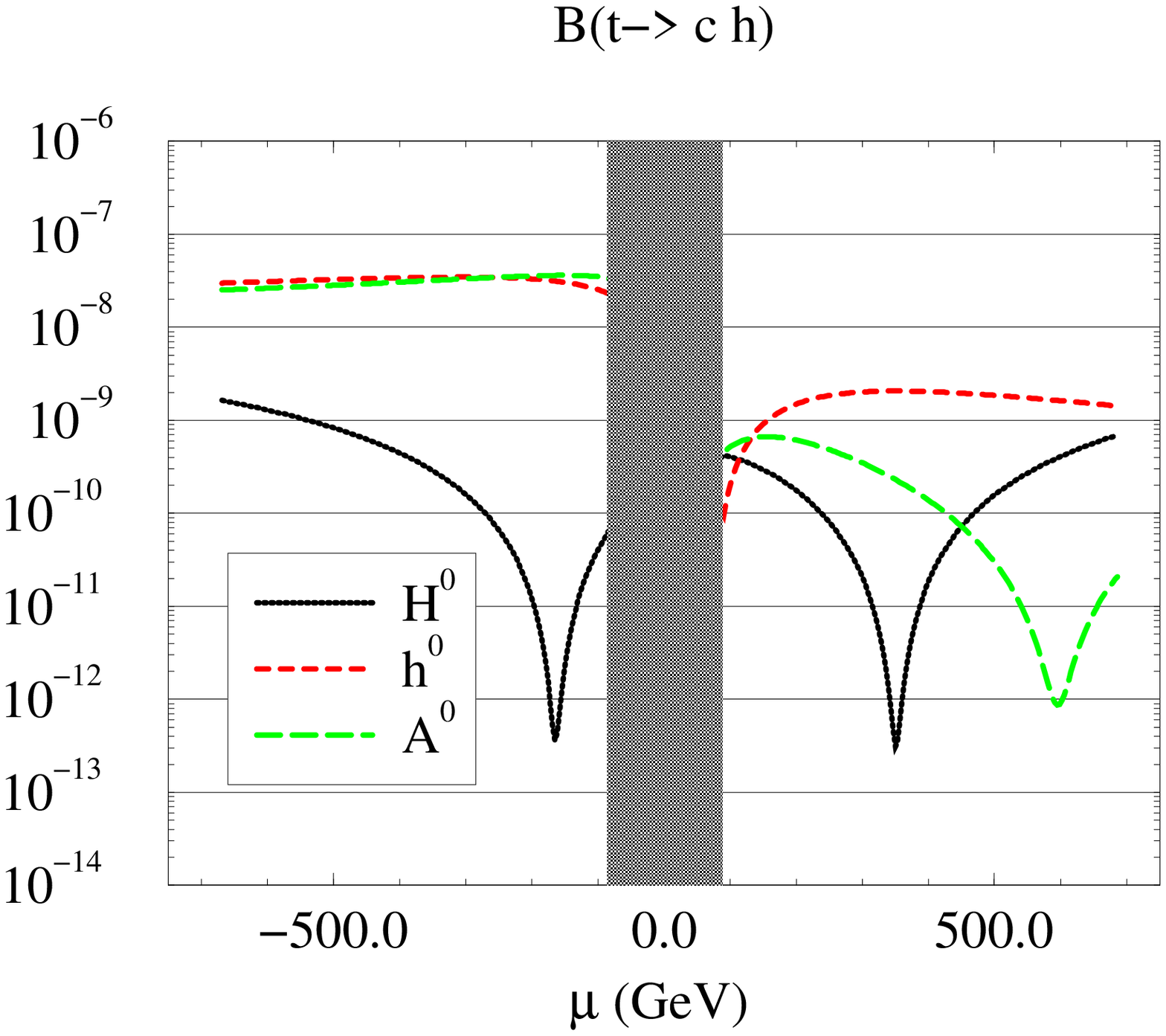,width=7cm}&
    \epsfig{file=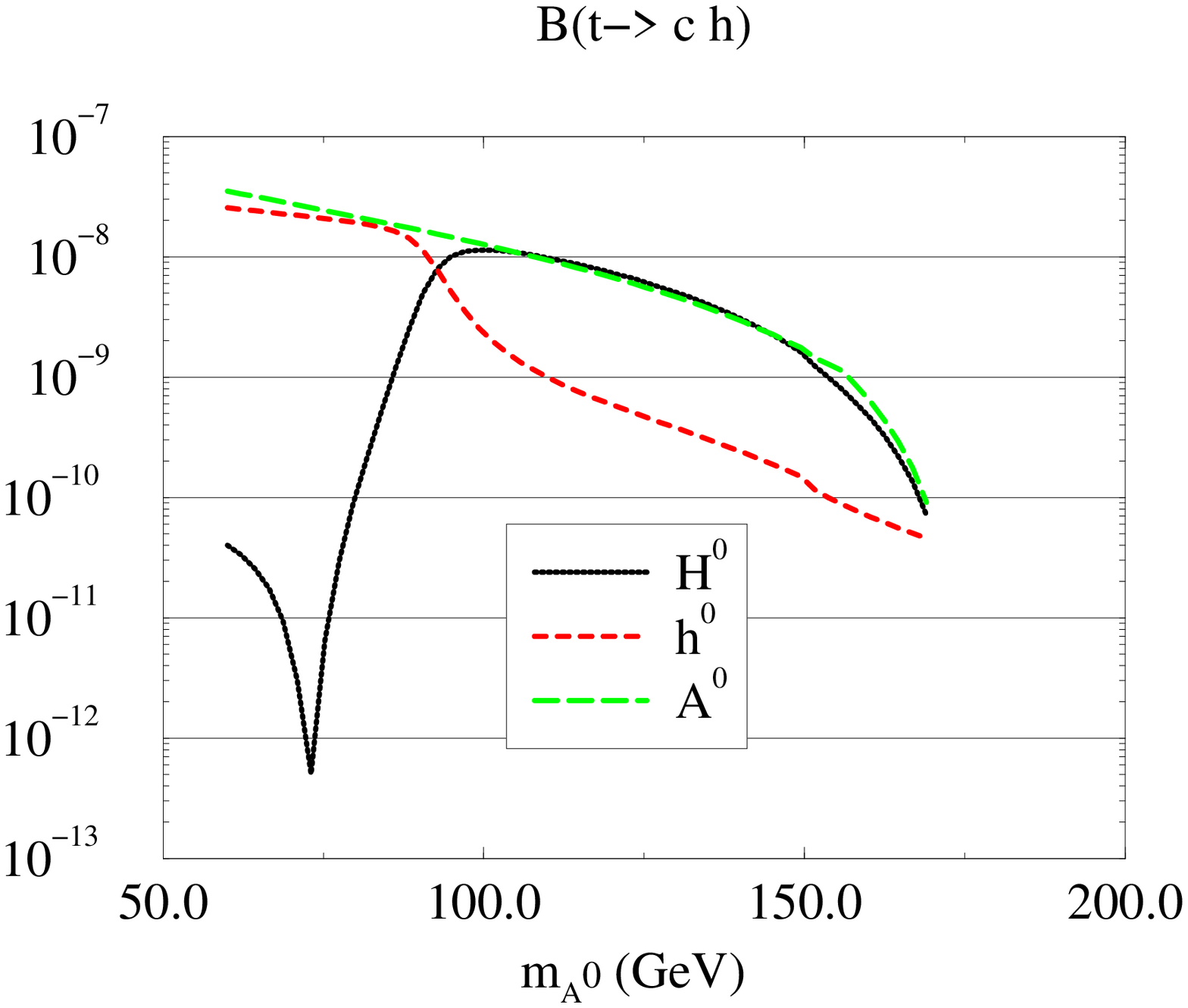,width=7cm}\\
    (c) & (d)
  \end{tabular}

\vspace{.3cm}

  {\Large Fig. 3}                      
\end{center}

\newpage

\begin{center}
  \begin{tabular}{cc}
    \epsfig{file=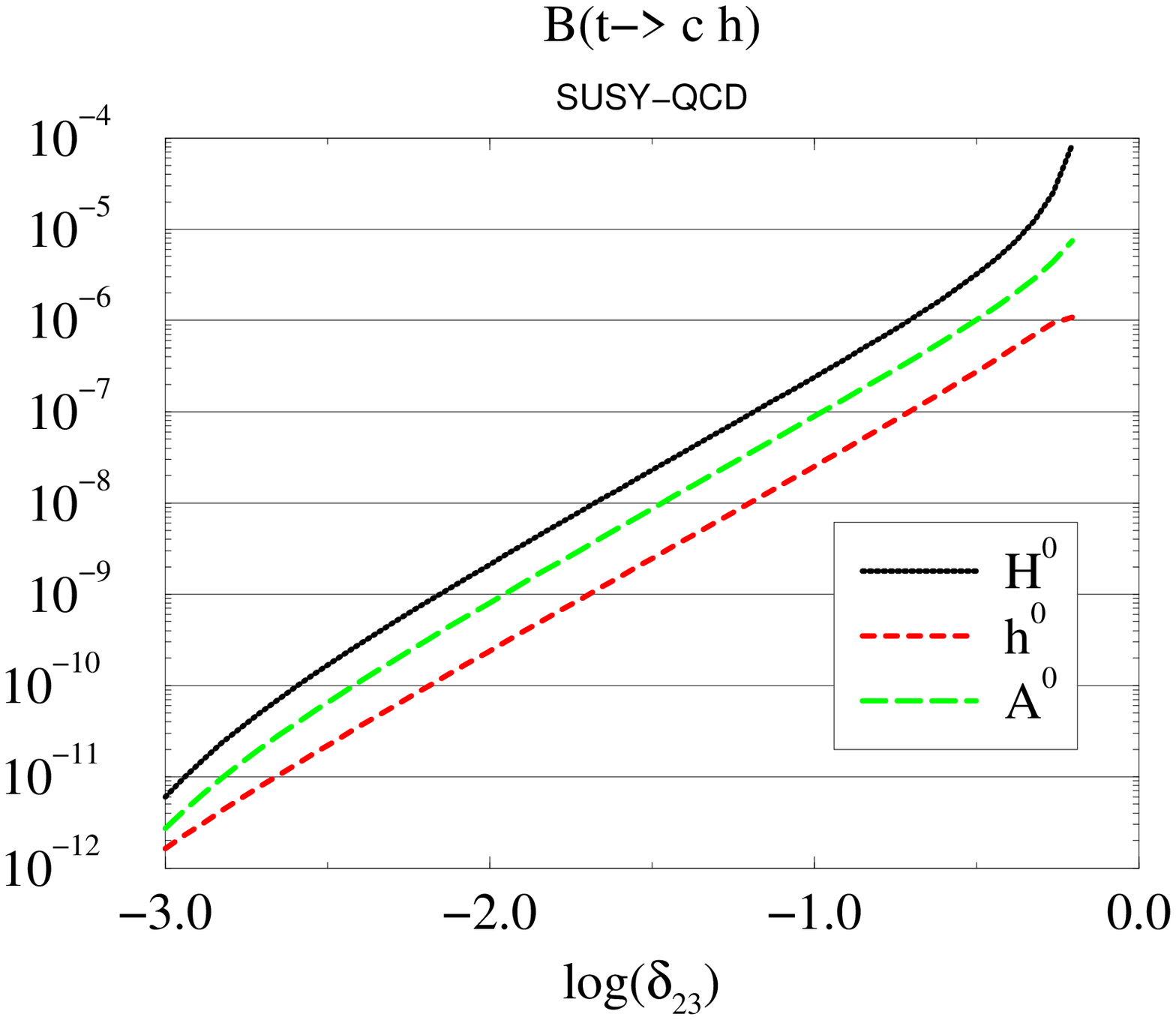,width=7cm} &
    \epsfig{file=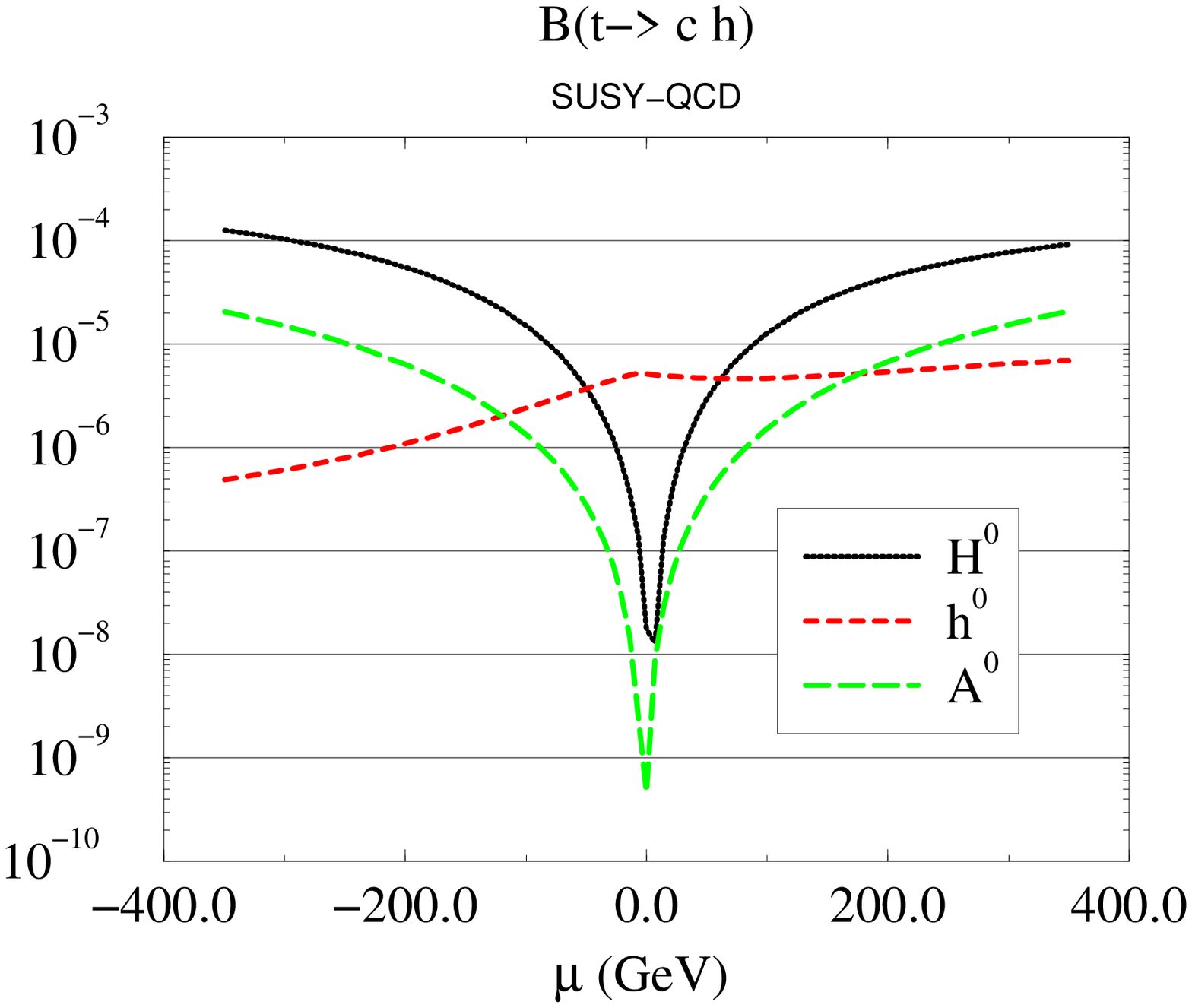,width=7cm}\\
    (a) & (b) \\
    \epsfig{file=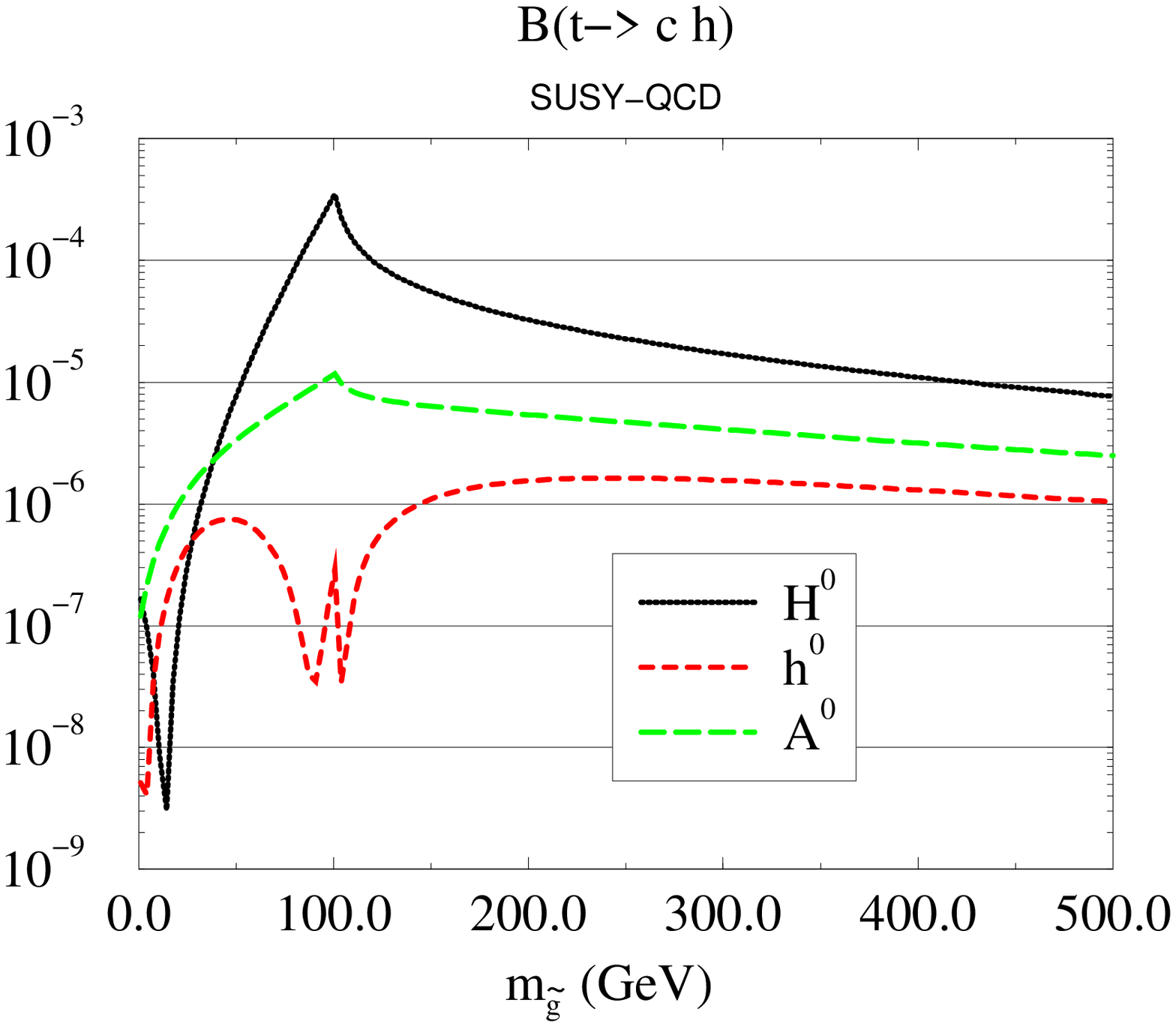,width=7cm}&
    \epsfig{file=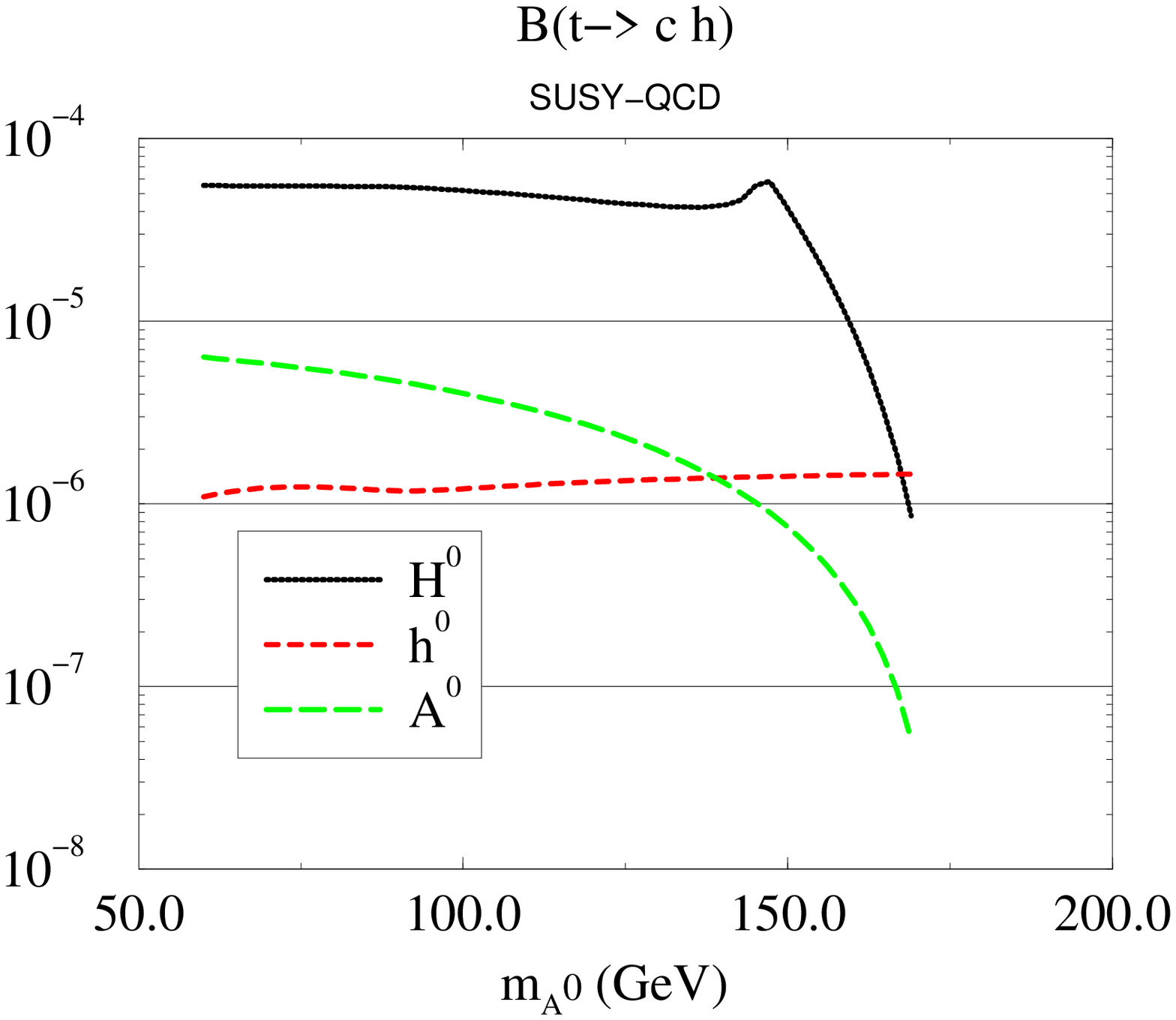,width=7cm}\\
    (c) & (d)
  \end{tabular}

\vspace{.3cm}

  {\Large Fig. 4}                      
\end{center}

\end{document}